\documentclass[sigconf]{acmart}

\usepackage[utf8]{inputenc}
\usepackage{booktabs}
\usepackage{natbib}
\usepackage{hyperref}
\usepackage{lineno} 
\usepackage{graphicx}
\usepackage{subcaption}
\usepackage{float}
\usepackage{enumitem}
\usepackage{multirow}
\usepackage{threeparttable}
\graphicspath{ {./figures/} }

\usepackage{tikz}
    \usetikzlibrary {positioning}
    \definecolor {processblue}{cmyk}{0.96,0,0,0}
    \usetikzlibrary{calc}
    \usetikzlibrary{shapes.geometric}
    \usetikzlibrary{backgrounds}
    \usepackage{varwidth}

\newcommand{\mask}{\underline{\hspace{0.75cm}}}
\newcommand{\sep}{\ $\Vert$\ }
\newcommand{\texta}{\ensuremath{R_1}}
\newcommand{\textb}{\ensuremath{R_2}}
\newcommand{\pattern}{P}
\newcommand{\patternarg}{\mathbf{x}}
\newcommand{\verbalizer}{v}
\newcommand{\pvp}{\mathbf{p}}

\newcommand{\trainset}{\mathcal{T}}
\newcommand{\unlabeledset}{\mathcal{D}}

\newcommand{\setLabel}{\mathcal{L}}
\newcommand{\indLabel}{\ell}
\newcommand{\lmvocabulary}{\mathcal{V}_\text{LM}}
\newcommand{\petensemble}{\mathcal{M}}
\newcommand{\indmodel}{m}
\newcommand{\indunlabeled}{d}
\DeclareMathOperator{\modelweight}{w}
\DeclareMathOperator{\score}{s}
\acmConference[CASCON'22]{32nd Annual International Conference on Computer Science and Software Engineering}{Nov 15--17 2022}{Toronto, Canada}
\setcopyright{rightsretained} 
\acmDOI{}
\acmISBN{}

\begin{document}\sloppy


\title[A Prompt-based Few-shot Learning Approach to Software Conflict Detection]{A Prompt-based Few-shot Learning Approach to Software Conflict Detection}

\author{Robert K. Helmeczi} 
\affiliation{
  \institution{Data Science Lab \\Ryerson University}
  \city{Toronto}
  \state{Ontario}
  \country{Canada}
  }
\email{rhelmeczi@ryerson.ca}

\author{Mucahit Cevik}
\affiliation{
  \institution{Data Science Lab \\Ryerson University}
  \city{Toronto}
  \state{Ontario}
  \country{Canada}
  }
\email{mcevik@ryerson.ca}

\author{Savas Yıldırım}
\affiliation{
  \institution{Data Science Lab \\Ryerson University}
  \city{Toronto}
  \state{Ontario}
  \country{Canada}
  }
\email{savas.yildirim@ryerson.ca}

\begin{abstract}
A software requirement specification (SRS) document is an essential part of the software development life cycle which outlines the requirements that a software program in development must satisfy. This document is often specified by a diverse group of stakeholders and is subject to continual change, making the process of maintaining the document and detecting conflicts between requirements an essential task in software development. Notably, projects that do not address conflicts in the SRS document early on face considerable problems later in the development life cycle. These problems incur substantial costs in terms of time and money, and these costs often become insurmountable barriers that ultimately result in the termination of a software project altogether. As a result, early detection of SRS conflicts is critical to project sustainability. The conflict detection task is approached in numerous ways, many of which require a significant amount of manual intervention from developers, or require access to a large amount of labeled, task-specific training data. In this work, we propose using a prompt-based learning approach to perform few-shot learning for conflict detection. We compare our results to supervised learning approaches that use pretrained language models, such as BERT and its variants. Our results show that prompting with just 32 labeled examples can achieve a similar level of performance in many key metrics to that of supervised learning on training sets that are magnitudes larger in size. In contrast to many other conflict detection approaches, we make no assumptions about the type of underlying requirements, allowing us to analyze pairings of both functional and non-functional requirements. This allows us to omit the potentially expensive task of filtering out non-functional requirements from our dataset.
\end{abstract}


\settopmatter{printfolios=true} 
\maketitle

\section{Introduction}

A software requirement specification (SRS) document outlines the desired behaviour of a new software program. These requirements often specify both functional and non-functional requirements. A coherent SRS is essential to the software development process, as conflicting requirements which go undetected can result in a considerable amount of lost time, particularly if the development process is late into the software development cycle~\citep{osman2018ambiguous, aldekhail2016software}. While some entities can absorb the cost associated with recovering from this time loss, issues with SRS documents can often result in the termination of a software development project altogether~\citep{aldekhail2016software}. As a result, ensuring that the requirement specifications do not conflict with each other is an essential task in software development. This task is made especially difficult for large software projects where numerous different parties can contribute to the specifications, and where requirements are continually subject to change~\citep{guo2021automatically, zowghi2002three}.

In this work, we consider a conflict detection task described as follows. Given an SRS document containing $N$ requirements, the number of pairings of requirements is $\mathcal{O}(N^2)$. The task is to assign one of the following labels to each pairing: \textit{Conflict}, indicating that the pairing contains two conflicting requirements; \textit{Duplicate}, indicating that the pairing contains two requirements that are equivalent; and \textit{Neutral}, indicating that the pairing contains two mutually independent requirements. Detecting duplicates is important in part because, if one requirement is to change, it must do so while still satisfying the requirements specified by its duplicates~\citep{guo2021automatically}. For particularly large software development projects with many requirements, it is infeasible to label each pairing manually. Instead, we propose to use few-shot learning, specifically through pattern-exploiting training (PET), to label each pairing~\citep{schick2021exploiting}. PET is a prompt-based learning method which leverages access to a large set of unlabeled pairings, $\unlabeledset$, and a small set of labeled pairings, $\trainset$. In the conflict detection domain where it is feasible to manually label a few requirement pairings, many of the remaining pairings to be labeled can be used as $\unlabeledset$.

PET is built upon pretrained language models (PLMs), such as BERT and its variants, but it is able to classify sequences with high accuracy while using far fewer labeled examples by rephrasing inputs as cloze-style questions. This rephrasing provides context to the task that is solved, in our case providing the PLM with knowledge of the conflict detection task.

The objective of this research is to establish PET as a viable candidate for few-shot learning for the conflict detection tasks. The contributions of our study can be summarized as follows:
\begin{itemize}\setlength\itemsep{0.45em}
    \item We extend \citet{Malik2022}'s work on conflict detection using transformer models to a few-shot setting. In particular, we provide a direct comparison of BERT and its variants on an array of training sets of size 32 up to 2,048. Our results detail the few shot performance of each transformer model and also quantify the impact of gradually increasing training set size on model performance.
    
    \item We investigate the impact of reformulating input examples as cloze-style phrases on few-shot performance. Specifically, we consider reformulations adopted from \citet{schick2021exploiting, schick2021fewshot} and introduce several of our own patterns. This investigation serves two purposes: firstly, it provides insight into how well generic reformulations transfer into the conflict detection task; and secondly, it provides a comparison of generic reformulations with ones that are targeted specifically towards conflict detection.
    
    \item We demonstrate the viability of few-shot learning for conflict detection. Specifically, we show that, given access to a small set of labeled data, prompt-based learning through pattern-exploiting training can produce models which are comparable in terms of well-known performance metrics to supervised learners trained on substantially larger training sets.
\end{itemize}

The rest of the paper is organized as follows. Section~\ref{sec:lit_rev} provides a review of the relevant studies in conflict detection and machine learning techniques for SRS text classification. Section~\ref{sec:methodology} provides an overview of our conflict detection dataset, transformer models, and the PET algorithm. Section~\ref{sec:results} summarizes the few-shot performance of the considered PLMs and demonstrates the few-shot abilities of PET for the conflict detection tasks. Section~\ref{sec:conclusion} summarizes our findings and proposes future research directions as extensions to our research.

\section{Literature Review}\label{sec:lit_rev}
SRS documents form the premise for a considerable amount of research due to their importance in the software development life cycle.
Research into SRS documents includes requirements classification~\citep{rahimi2020ensemble, dias2020software}, ambiguity detection~\citep{osman2018ambiguous, umber2011minimizing}, and fault detection~\citep{alshazly2014defect, thelin2003experimental}.
Many studies also focus on introducing various natural language processing (NLP) techniques into the software requirements domain~\citep{nazir2017applications, kici2021text, dias2020software, umber2011minimizing, elhassan2022requirements}.

\citet{kici2021text} investigate text classification techniques for SRS documents using transfer learning through transformer models. In particular, they investigate the performance of BERT, DistillBERT, RoBERTa, ALBERT, and XLNet on several different requirements datasets, some of which contain both functional and non-functional requirements, and several of which contain less than 1,000 total samples. 
In our analysis, we extend \citet{kici2021text}'s comparative analysis by investigating the impact of training set size on the performance of transformer models. 
While~\citet{kici2021text} achieve strong performance across several of their datasets, as their models are specifically fine-tuned on each classification task, their investigation assumes that a substantially large number of requirements for a particular project are already labeled. Our application of the PET algorithm offers a means for training a model using considerably fewer labeled instances than an equally-performant supervised learner.

Conflict detection in SRS documents is a popular research topic~\citep{yang2005process, sardinha2013ea, guo2021automatically, aldekhail2016software}. \citet{yang2005process} develop RECOMA (REquirements COnflicts MAnagement tool) for both identifying and managing the conflict detection process. Notably, their research attempts to provide a means for conflict detection while acknowledging that formal techniques for developing SRS documents are often infeasible due to the diverse set of stakeholders, many of whom cannot be expected to understand the language that formal methods require~\citep{guo2021automatically, yang2005process}. RECOMA relies on a data preprocessing step by proposing a rigid structure for outlining requirements by splitting requirements into an object, a verb, and a resource. This preprocessing is often delegated to a developer, who must then translate requirements to this structure before conflicts can be detected in RECOMA. Additionally, the RECOMA approach only extends to functional requirements.

\citet{guo2021automatically} propose FSARC (finer semantic analysis-based requirements detector), an automated approach to conflict detection for functional requirements front-lined by semantic analysis of documents. Their approach requires first converting requirements to an eight-tuple that forms a harmonized semantic meta-model~\citep{guo2021automatically}. Each semantic element in a requirement is identified using a separate algorithm. The identification of semantic elements relies on heuristics and requires the manual labeling of a small number of requirements, where labels can often be ambiguous resulting in discrepancies between multiple human labelers~\citep{guo2021automatically}. \citet{guo2021automatically} present promising results for FSARC, indicating that in spite of potential discrepancies between human labelers, the algorithm can still perform well in many circumstances. However, as this algorithm relies on converting requirements to an eight-tuple, in cases where identifying the tuple components is impossible or considerably ambiguous, the performance of the labeling algorithm may suffer considerably.

\citet{Malik2022} propose transformer models for conflict detection based on requirement pairings, and show that this approach can achieve significant performance especially when a large number of labeled pairings are available.
In this paper, we adapt \citet{Malik2022}'s approach to the conflict detection task in a few-shot setting.

To the best of our knowledge, no prompt-based method has been used in SRS conflict detection and, further still, few-shot learning has also not been applied to this task. Prompt-based learning is a relatively new paradigm in the NLP field. The GPT~\citep{gpt} and T5~\citep{t5} models are the strongest early examples of prompt-based learning. The GPT-3~\citep{gpt3} model achieves remarkable few-shot performance based on in-context learning by leveraging a natural-language prompt and a few task demonstrations. T5 shows that any NLP problem can be cast as text-to-text, which has been a major breakthrough in this field. Likewise, the autoencoder models reformulate downstream tasks with a masked language model (MLM) objective. Reformulating is done by adding task-specific tokens to the input sequence for conditioning, which gives us the ability to solve many problems by simply manipulating the input. With prompting, we can even train a model with no access to labeled data as we can directly rely on the objective function (i.e., the MLM objective)~\citep{schick2021exploiting}. 

Few-shot learning offers an advantage in conflict detection because it allows using task-specific labelings, particularly labelings for a specific SRS project. Specifically, the prompt-based learning approach that is explored in this study, PET, is able to predict conflicts with high precision using only a few labeled examples, allowing those involved in interpreting and implementing the requirements to spend considerably less time labeling the instances. Additionally, while PET is shown to be useful for few-shot learning, it offers a considerable performance boost even for larger training sets, particularly for difficult-to-label classes. This is despite the fact that the PLMs that are employed by PET are not trained on domain-specific data. Finally, we note that unlike some of the previously mentioned studies, PET does not assume the category of underlying requirements and should be able to label any pairing, including both functional and non-functional requirements. This introduces a particularly powerful advantage over several of the aforementioned studies as it ensures that requirements need not be filtered out based on their type before searching for conflicts.

\section{Methods}\label{sec:methodology}

In this section we provide a brief overview of the conflict detection dataset, describe the PLMs used in our experiments, and discuss pattern-exploiting training (PET).

\subsection{Conflict detection dataset}

We employ a proprietary conflict detection dataset obtained from \citet{Malik2022}'s work. Given a requirement pairing $\patternarg=(\texta{}, \textb{})$ from this dataset, the available labels are \textit{Conflict}; \textit{Duplicate}; and \textit{Neutral}, indicating that the requirements are in disagreement; agreement; or are mutually independent of one another. Table~\ref{tab:conflict-detection-examples} outlines an example for each available label and it also helps demonstrate the difficulties associated with this task. In particular, while we expect it to be relatively easy to detect neutral pairings, the vocabulary used between conflicting and duplicated requirements tends to be very similar. 

\begin{table*}[htb]
\caption{Conflict detection examples.}
\label{tab:conflict-detection-examples}
\resizebox{0.85\textwidth}{!}{
\begin{tabular}{p{0.35\linewidth}p{0.35\linewidth}l}
\toprule
Specification 1 & Specification 2 & Label \\
\midrule
The UAV shall instantaneously transmit information to the Pilot regarding mission-impacting failures. & The Hummingbird shall send the Pilot real-time information about malfunctions that impact the mission. & Duplicate \\
\midrule
The UAV shall only accept commands from an authenticated Pilot. & The UAV shall accept commands from any Pilot controller. & Conflict \\
\midrule
The UAV flight range shall be at least 20 miles from origin. & The UAV shall be able to transmit video feed to the Pilot and up to 4 separate UAV Viewer devices at once. & Neutral \\
\bottomrule
\end{tabular}
}
\end{table*}

PET is able to perform few-shot learning in part through the use of a large set of unlabeled requirement pairings, $\unlabeledset$. To simulate this unlabeled set, we took 5,000 labeled requirement pairings and discarded their labels.
The conflict dataset that we consider is relatively small, containing about 10,000 requirement pairings which are then split between the training, unlabeled, and test sets. In comparison, \citet{schick2021exploiting} typically considered problems where $\unlabeledset$ alone had at least 20,000 examples. To split our data, we performed stratified sampling for each of the training, unlabeled, and test sets. The distribution of class labels for each set is outlined in Table~\ref{tab:conflict-dataset-summary}.

\setlength{\tabcolsep}{4pt} 
\renewcommand{\arraystretch}{1.3} 
\begin{table}[htb]
\caption{Conflict detection class distribution.}
\label{tab:conflict-dataset-summary}
\begin{threeparttable} 
    \begin{tabular}{lrrr}
    \toprule
              &  Conflict &  Duplicate &  Neutral \\
    \midrule
    $|\trainset| = 32$ &        17 &          5 &       10 \\
    $|\trainset| = 64$ &        33 &         10 &       21 \\
    $|\trainset| = 128$ &        67 &         20 &       41 \\
    $|\trainset| = 256$ &       134 &         40 &       82 \\
    $|\trainset| = 512$ &       267 &         81 &      164 \\
    $|\trainset| = 1{,}024$ &       535 &        161 &      328 \\
    $|\trainset| = 2{,}048$ &      1,070 &        323 &      655 \\
    $\unlabeledset$ (unlabeled$^\dagger$) &      2,613 &        787 &     1,600 \\
    Test      &      1,045 &        315 &      640 \\
    \bottomrule
    \end{tabular}
    \begin{tablenotes}
        \item[$\dagger$] Labels are removed for training.
    \end{tablenotes}
\end{threeparttable}
\end{table}

\subsection{Transformer models}

In our analysis, we deploy several PLM checkpoints, specifically uncased BERT, uncased BERT-large, RoBERTa, RoBERTa-large, ALBERT-v2, and ALBERT-xxlarge-v2.
Each PLM is trained on a large corpus of unlabeled data, and is then fine-tuned on a downstream task~\citep{devlin2019bert}.
We first evaluate the performance of sequence classification using these PLMs to determine the best performer on our dataset. We then choose the best performing PLM for our prompt-based learning task.
\begin{itemize}\setlength\itemsep{0.45em}
    \item \textbf{BERT (Bidirectional Encoder Representations from Transformers)} models are trained on case insensitive data. These are the base version (BERT-uncased) which has a total of 110M parameters, and the large version (BERT-large-uncased) which has 340M parameters~\citep{devlin2019bert}. BERT employs bidirectional self-attention, allowing the model to attend to both previous and future tokens in the self-attention layer~\citep{devlin2019bert}. This bidirectional training is made possible through the use of a masked language model~\citep{devlin2019bert}. Specifically, in the pretraining objective, tokens are randomly masked and the model must predict the appropriate token based on its context.
    
    \item \textbf{RoBERTa (Robustly optimized BERT approach)} is a modified version of BERT which implements the changes outlined by~\citet{liu2019roberta}. These modifications include increasing the batch size, training for more steps, using a larger pretrain dataset, and employing a more generalized text encoding. 
    In the MLM objective used for BERT pretraining,  masking is done as a pre-processing step, which results in the same masked sequences being passed to the model~\citep{liu2019roberta}. In contrast, RoBERTa applies masking in a dynamic fashion, generating the masking pattern for each sequence as it is passed to the model~\citep{liu2019roberta}. Following from the BERT models, RoBERTa-base corresponds to BERT-base and RoBERTa-large corresponds to BERT-large. The modified text encoding adds parameters to RoBERTa, with RoBERTa-base using 125M parameters and RoBERTa-large using 360M parameters.
    
    \item \textbf{ALBERT (A Lite BERT)} is another BERT variant, this time designed to improve training times and lower memory usage. ALBERT introduces two major techniques to allow for considerably improved scalability. Notably, ALBERT-base-v2 has just 12M parameters compared to the 110M in BERT-base, and ALBERT-xxlarge-v2 has 235M parameters, less than the 340M in BERT-large~\citep{lan2020albert}. The original ALBERT proposed in~\citet{lan2020albert} is later improved by adjusting several training parameters\footnote{\url{https://github.com/google-research/albert}}. We employ this updated version, denoted by ``v2'', in our work.
\end{itemize}

In our supervised learning experiments, we employ the sequence pair classification ability of each of the aforementioned language models to label our data. 

\subsection{Pattern-exploiting training}

PET is a prompt-based learning approach introduced by~\citet{schick2021exploiting} for few-shot learning tasks.
The contribution of PET to the labeling process is to provide context to the task being solved.
Few-shot learning is a method for classifying instances with access to only a small number of labeled examples.
This is in contrast to standard supervised learning techniques, which assume access to a large number of labeled training instances.
Consider a simplified version of the conflict detection task, where our goal is to label a sentence pair $\patternarg=(\texta, \textb)$, as either \textit{Duplicate} or \textit{Not Duplicate}.
In a few-shot setting, this is a difficult problem to solve for a standard sequence classifier as it has no context about the task at hand, simply seeing the two sequences.
In PET, we provide context by mapping the two requirements to a cloze-style phrase (i.e., we introduce a ``prompt'').
In this context, we might ask the question:
\begin{quote}
    Does ``$\texta$'' imply ``$\textb$''? \mask{}.
\end{quote}
where the allowed answers for the masked token ``\mask{}'' in this context are \textit{Yes} for \textit{Duplicate}, and \textit{No} for \textit{Not Duplicate}. In this case, we can use the PLMs with an MLM objective to predict the correct labels.

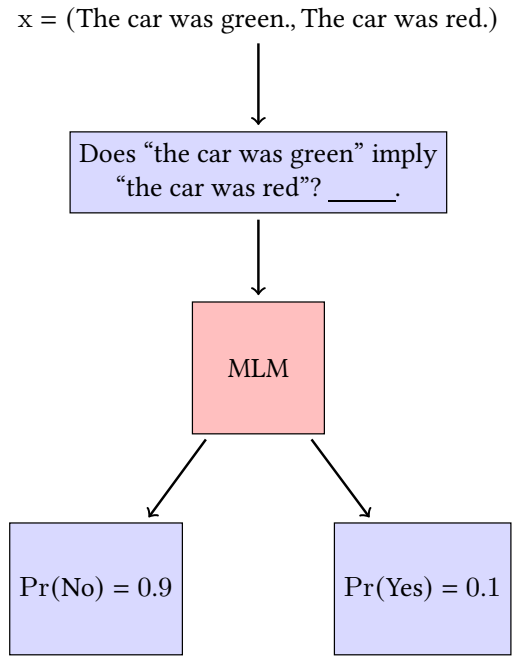
\begin{figure}[!ht]
    \centering
    \resizebox{0.8\linewidth}{!}{\begin {tikzpicture}[node distance =1 cm and 2cm]
    \node (patternarg) {\begin{varwidth}{20em}\centering $\patternarg=(\text{The car was green.}, \text{The car was red.})$\end{varwidth}};
    \node[draw, fill=blue!15] [below=of patternarg](pattern) {\begin{varwidth}{15em}\centering Does ``the car was green'' imply\\ ``the car was red''? \mask{}.\end{varwidth}};
    \node[draw, fill=pink, minimum size=1.5cm] [below=of pattern](mlm) {\begin{varwidth}{15em}\centering MLM\end{varwidth}};
    \draw[->, thick, shorten >= 2pt, shorten <= 0pt] (patternarg) edge node {} (pattern);
    \draw[->, thick, shorten >= 2pt, shorten <= 2pt] (pattern) edge node {} (mlm);
    \node[draw, fill=blue!15, minimum size=1.5cm] [below left=1cm and 0.1cm of mlm](no) {\begin{varwidth}{15em}\centering $\Pr(\text{No})=0.9$\end{varwidth}};
    \node[draw, fill=blue!15, minimum size=1.5cm] [below right=1cm and 0.1cm of mlm](yes) {\begin{varwidth}{15em}\centering $\Pr(\text{Yes})=0.1$\end{varwidth}};
    \draw[->, thick, shorten >= 2pt, shorten <= 2pt] (mlm) edge node {} (no);
    \draw[->, thick, shorten >= 2pt, shorten <= 2pt] (mlm) edge node {} (yes);
    \end{tikzpicture}
    }
    \caption{Example application of PET on an input example.}
    \label{fig:mlm-predict-sentence}
\end{figure}

Figure~\ref{fig:mlm-predict-sentence} illustrates an application of the aforementioned context-enriching scenario.
In this figure, the objective is to classify the pairing $\patternarg=(\textit{The car was green.}, \textit{The car was red.})$ as either \textit{Duplicate} or \textit{Not Duplicate}.
Notably, these two statements do not immediately fit into the pattern structure: we must perform some preprocessing to remove the terminating punctuation and the initial capital letter.
Once this is done, the pattern with the substituted argument $\patternarg$ is passed to the MLM.
The MLM returns the probability that the masked token ``\mask{}'' should be \textit{No} and the probability that it should be \textit{Yes}, where \textit{No} indicates that the two statements should be marked \textit{Not Duplicate}, and \textit{Yes} indicates that they should be marked \textit{Duplicate}.
These probabilities can then immediately be converted to the label probabilities, and as \textit{No} has the higher probability here, this means the label for this input example is predicted to be \textit{Not Duplicate}.

Formalizing this example, we say that a \textit{pattern} $\pattern$ is a mapping of an input $\patternarg$ to a cloze-style phrase that contains a masked token~\citep{schick2021exploiting}, which,
in the case of conflict detection, can be taken as $\patternarg=(\texta{}, \textb{})$. 
Additionally, we define a verbalizer $\verbalizer$ as a bijective function which maps each label $\indLabel\in\setLabel$ to some token or sequence of tokens in the vocabulary of the PLM, $\lmvocabulary$~\citep{schick2021exploiting}. 
We can then employ a MLM to predict the most likely substitute for the masked token, and apply $\verbalizer^{-1}$ to recover the true label. 
We define a pattern-verbalizer pair (PVP) as $\pvp=(\pattern, \verbalizer)$ (i.e., it is a tuple of a pattern and a verbalizer).

In theory, PET is able to improve the performance of a learner by adding context to the target problem. In practice however, this is made difficult by several factors. Firstly, in a few-shot setting there is no available data to fine-tune PVPs, allowing for some PVPs to perform worse than sequence classification alone. Secondly, every PVP requires training a separate model, increasing the time required for a PET ensemble to predict the label for a given example $\patternarg$. In the presence of a relatively large set of unlabeled data instances, these issues can be addressed in part through the use of knowledge distillation as outlined by~\citet{hinton2015distilling}.

Given a set of unlabeled data instances $\unlabeledset$ and an ensemble of PET models $\petensemble$, we can use each $\indmodel\in\petensemble$ to generate soft labels, or scores, for each $\indunlabeled\in\unlabeledset$~\citep{schick2021exploiting}. These scores are then combined using a weighted sum where the weights are the accuracy of each PVP before training~\citep{schick2021exploiting}. By using the weight before training, those models which more naturally describe the task even without access to labeled data are assumed to employ superior PVPs. Given the final soft labels, a sequence classifier is trained on the training set together with the softly labeled set $\unlabeledset$.

When a MLM $\indmodel$ predicts a substitute for the masked token, it assigns a score to each of the tokens in its vocabulary. For our purposes, the important scores are the ones associated with the given verbalizations. In this case, the score assigned to the label $\indLabel$ by model $\indmodel$ given the pattern $\pattern(\patternarg)$ is $\score(\ell|\patternarg,\indmodel)$, and it is simply the score that the MLM assigns to $\verbalizer(\indLabel)$. After we train an ensemble of models $\petensemble$, we can combine the scores for each label $\indLabel\in\setLabel$ using Equation~\ref{eq:ensemble-score}~\citep{schick2021exploiting}:
\begin{align}
    \score(\ell|\patternarg,\petensemble) = \dfrac{\displaystyle{\sum_{\indmodel\in\petensemble}}\modelweight(\indmodel)\score(\ell|\patternarg, \indmodel)} {\displaystyle{\sum_{\indmodel\in\petensemble}}\modelweight(\indmodel)}
    \label{eq:ensemble-score}
\end{align}
where $\modelweight(\indmodel)$ denotes the weight assigned to model $\indmodel$. Finally, we convert label scores to probabilities using softmax.

Our observations, consistent with those in \citet{schick2021exploiting}, show that the performance of a PLM trained on a PVP can vary considerably between runs. As a result, for each PVP we train three separate models. When evaluating PVPs alone we report the average performance of these runs. In our final PET ensemble, we use all three iterations for each PVP selected. This means that for a PET ensemble of three PVPs, there are nine total models contributing to the predictions.

\subsection{Patterns}

In our investigation, we consider six different patterns: two as proposed by~\citet{schick2021exploiting, schick2021fewshot} (marked with an asterisk in the table), and four of our own. We list all patterns in Table~\ref{tab:patterns}. We mark our masked token(s) with a single ``\mask''. Notably, when a verbalizer consists of more than a single token, we must use multiple consecutive masked tokens as described by~\citet{schick2021fewshot}. In our experiments, all verbalizations require the use of two masked tokens. Additionally, we separate text segments with vertical bars ($\|$). This separation is handled differently by each PLM~\citep{schick2021exploiting}.
\begin{table*}[!h]
    \caption{PET patterns for conflict detection.}
    \label{tab:patterns}
    \resizebox{0.85\linewidth}{!}{
        \begin{tabular}{lp{0.75\linewidth}}
        \toprule
        ID & Pattern \\
        \midrule
        $\pattern_1^*(\patternarg)$ & ``\texta{}''? \sep{} \mask{}, ``\textb{}''\\
        $\pattern_2^*(\patternarg)$ & \texta{}? \sep{} \mask{}, \textb{}\\
        $\pattern_3(\patternarg)$ & Given ``\texta{}'', we can conclude that ``\textb{}'' is \mask{}.\\
        $\pattern_4(\patternarg)$ & ``\texta{}'' means ``\textb{}''. \sep{} \mask{}.\\
        $\pattern_5(\patternarg)$ & ``\texta{}'' implies ``\textb{}'' and ``\textb{}'' implies ``\texta{}''.\sep{} The previous sentence is correct: \mask{}.\\
        $\pattern_6(\patternarg)$ & ``\texta{}'' is equivalent to ``\textb{}''. Similarly, ``\textb{}'' is equivalent to ``\texta{}''. \sep{} The previous statements are \mask{}. \\
        \bottomrule
    \end{tabular}
    }
\end{table*}

Observe that both of the patterns adopted from~\citet{schick2021exploiting, schick2021fewshot} do not include a trailing punctuation mark on \textb{}. Instead, these patterns use the punctuation already available in \textb{}. In contrast, all of our patterns strip the trailing punctuation from both \texta{} and \textb{}. Additionally, we quote the statements that are input into the pattern, ensuring a natural transition between the input and the rest of the pattern.

The patterns provided by~\citet{schick2021exploiting, schick2021fewshot} are basic, contributing only punctuation and a separator. In contrast, we develop patterns which attempt to facilitate the input $\patternarg$ as a natural extension of the text. Patterns 5 and 6 additionally attempt to embed the commutative property of the requirements $\texta$ and $\textb$: the label is unaffected when the two sequences are swapped. As PLMs are trained on a finite sequence length, these ``commutative patterns'' are not always possible. However, for the conflict detection dataset with a sequence length of 256, we found that we never exceeded this maximum sequence length. 

\subsection{Verbalizers}

As with developing patterns, the selection of verbalizers can have a considerable impact on the performance. 
Without access to a development set, it is difficult to evaluate how well a verbalizer will perform. \citet{schick2020petal} discuss a method of automatically choosing verbalizers. This may be especially useful for high cardinality tasks where it is difficult to choose any verbalization other than the identity, which maps a label to itself. Notably, however, their results showed that hand picked verbalizers tend to outperform the automatically generated ones. As a result, we develop our own patterns and verbalizers for our classification tasks. They noted that PLMs have a bias towards frequent words, making verbalizations of frequent words superior to verbalizations with rare words~\citep{schick2020petal}. Accordingly, we attempt to use common words in our verbalization.

Table~\ref{tab:verbalizers} lists the verbalizers that we used for this task. We consider one verbalizer provided by \citet{schick2021exploiting}, marked with an asterisk, and introduce one of our own. Unlike with pattern design, the choice of verbalizer can have a considerable impact on runtime. The time taken to train a model is linear with respect to the number of tokens used in the verbalization. Here we consider verbalizers with a maximum of two masked tokens. 

\begin{table}[!htb]
    \centering
    \caption{Verbalizers.}
    \label{tab:verbalizers}
    \begin{tabular}{lccc}
    \toprule
        & $\verbalizer(\text{Conflict})$ & $\verbalizer(\text{Duplicate})$ & $\verbalizer(\text{Neutral})$  \\
    \midrule
        $\verbalizer_1^*$ & No & Yes & Maybe \\
        $\verbalizer_2$ & False & True & Neither \\
    \bottomrule
    \end{tabular}
\end{table}

\section{Results}\label{sec:results}

In our analysis, we consider labeled training sets sized in powers of two from $2^5=32$ up to $2^{11}=2{,}048$. We first investigate the performance of six PLMs: BERT and five of its variants. Then, we choose the best performing PLM to serve as the underlying MLM for our PET patterns. We investigate the performance of several PVPs by training three MLMs for each PVP and reporting the average performance on the test set. Finally, we train a PET ensemble consisting of three PVPs and compare its performance to that of the best supervised learner. 

\subsection{Hyperparameter selection}

Our selection of hyperparameters are largely derived from~\citet{schick2021exploiting, schick2021fewshot}. We note that in a few-shot learning setting it is unlikely that a development set for fine-tuning hyperparameters will be available. As a result, and in consensus with \citet{schick2021exploiting}'s suggestion, we choose hyperparameters from previous work, adopting the majority of our hyperparameters from~\citet{schick2021exploiting, schick2021fewshot}, with one adjustment. As per the findings in~\citet{zhang2021revisiting}'s study, which note that training for more batches tends to improve performance, we choose to increase the number of training steps from 250 to 1,000 for all learners except for the final sequence classifier trained on the PET ensemble. As this classifier is trained on the soft-labeled dataset $\unlabeledset$ which has 5,000 examples, we use 5,000 training steps, as in~\citep{schick2021exploiting}. Note that, as in~\citet{schick2021exploiting} the number of unlabeled examples was more than 20,000 for each task, training for 5,000 steps on our unlabeled set of size 5,000 represents an effective increase in the number of training steps, which is consistent with our adherence to the observations in~\citet{zhang2021revisiting}. We summarize the hyperparameters used, borrowing the programmatic variable names from the PET repository\footnote{\url{https://github.com/timoschick/pet}.}, in Table~\ref{tab:pet-params}.
\begin{table}[!ht]
    \centering
    \caption{Hyperparameters for PET.}
    \label{tab:pet-params}
    \begin{threeparttable}
    \begin{tabular}{lc}
\toprule
    Parameter & Value  \\
\midrule
    \texttt{max\_steps} & $1{,}000^{\dagger}$ \\
    \texttt{gradient\_accumulation\_steps} & 4 \\
    \texttt{learning\_rate} & 1e-5 \\
    \texttt{adam\_epsilon} & 1e-8 \\
    \texttt{warmup\_steps} & 0 \\
    \texttt{max\_grad\_norm} & 1.0 \\
    \texttt{max\_seq\_length} & 256 \\
    \texttt{temperature} & 2 \\
\bottomrule
\end{tabular}
\begin{tablenotes}
    \item[$\dagger$] For training the PET ensemble on $\trainset\cup\unlabeledset$, we use 5,000 steps as per~\citep{schick2021exploiting}.
\end{tablenotes}
\end{threeparttable}
\end{table}

\subsection{Supervised learning performance}

We first consider the performance of six transformer models---BERT-uncased, BERT-large-uncased, RoBERTa, RoBERTa-large, ALBERT-v2, and ALBERT-xxlarge-v2---on the conflict detection dataset. As PET is an application of these models, we choose the best performing one as the MLM for the PET algorithm. Notably, this experiment evaluates the few-shot performance of these transformer models when they are applied to a sequence classification task.

\begin{figure}[!ht]
    \centering
    \includegraphics[width=\linewidth]{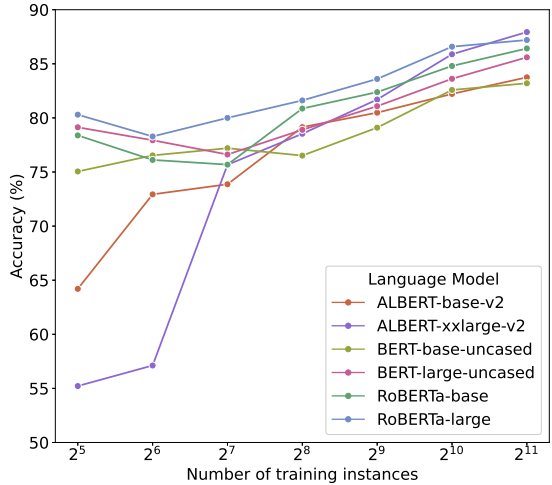}
    \caption{Supervised learning for conflict detection.}
    \label{fig:conflict-supervised-accuracy}
\end{figure}

Due to variance in the training process, we repeat training three times and report the average results. Figure~\ref{fig:conflict-supervised-accuracy} displays the performance of each language model on each training set. The results for $|\trainset|=2{,}048$, particularly the relative ordering of model performance, are similar to what we would expect from each model if they were trained on large training sets~\citep{devlin2019bert, liu2019roberta, lan2020albert}. However, this ordering is considerably different for small $|\trainset|$. As a consequence, we cannot assume that the performance of a PLM for few-shot learning will reflect its performance on full-sized training sets, as was done in~\citet{schick2021fewshot}. Additionally, we observe a more than 25\% difference between the best and worst performers when using 32 training instances. As PET is built on-top of a selected PLM, this performance impact would likely be considerable when using the PET algorithm as well.

We found that RoBERTa-large was consistently the best performing model, except for $|\trainset|=2{,}048$ where it performs moderately worse than ALBERT-xxlarge-v2. As a result, we choose RoBERTa-large as our baseline supervised learner to which we compare our PET results. Additionally, RoBERTa-large is chosen as the underlying MLM for the PET.

\subsection{PVP evaluation}

\begin{figure}[!ht]
    \centering
    \includegraphics[width=\linewidth]{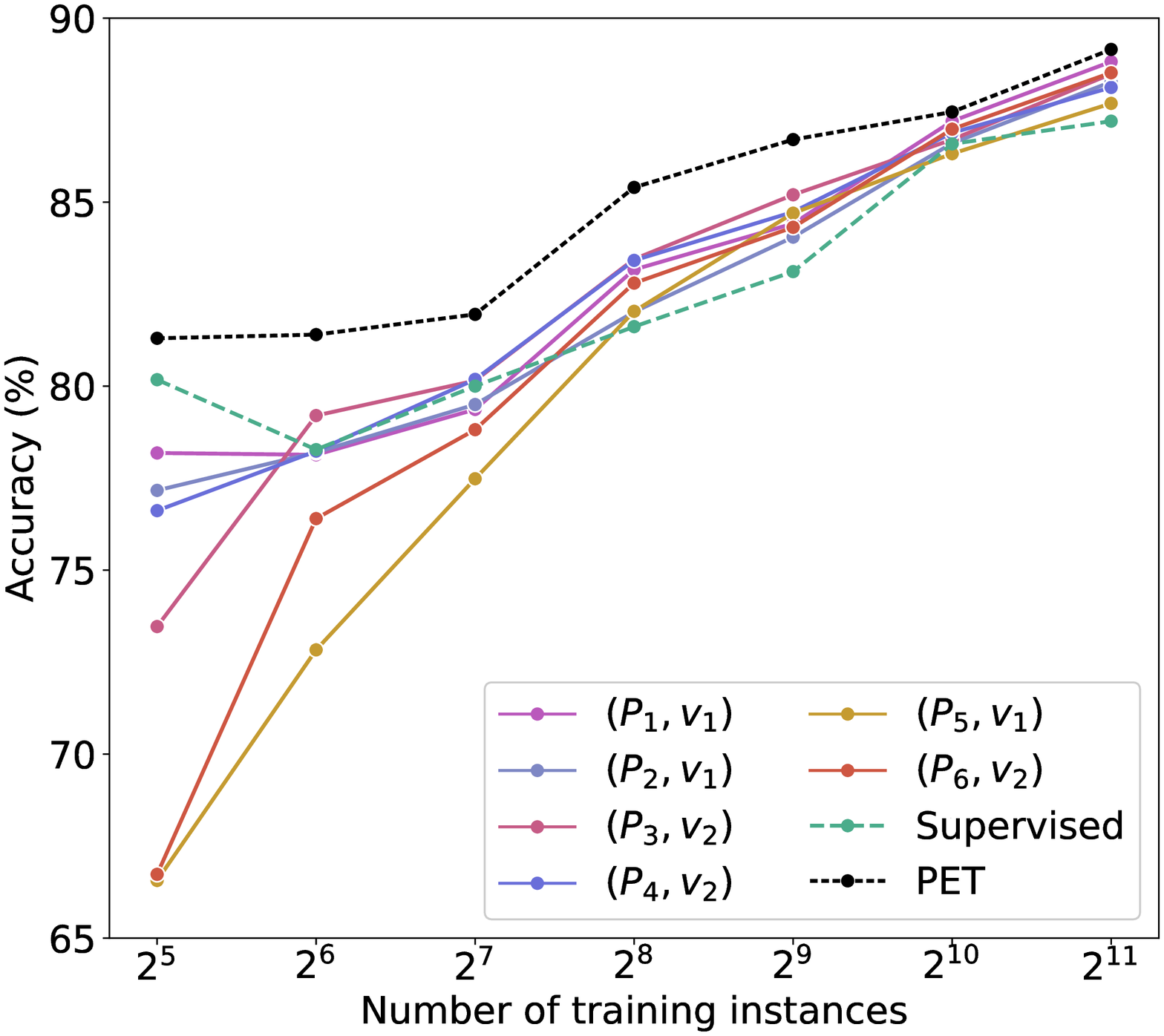}
    \caption{PET for conflict detection.}
    \label{fig:conflict-patterns}
\end{figure}

In addition to the patterns and verbalizers introduced by~\citet{schick2021exploiting, schick2021fewshot}, we developed our own in an attempt to more naturally describe the conflict detection task. However, Figure~\ref{fig:conflict-patterns} shows that the PVPs that we designed were no more competitive than those that we adopted, and in some cases ours performed even worse. We also observe that many of the PVPs fail to match up to traditional supervised learning, particularly for smaller training set sizes. These results are welcome as they suggest that patterns, like other hyperparameters, can often be borrowed from previous work. Specifically, while the PVPs introduced by~\citet{schick2021exploiting,schick2021fewshot} were not targeted to a conflict detection task, they transfer very well into this task with no modifications.

\subsection{Pattern-exploiting training performance}

For our PET ensemble, we consider three PVPs: $(P_1, v_1)$, $(P_3, v_2)$, and $(P_4, v_2)$. Due to random variations when training, we follow \citet{schick2021exploiting} and train three language models for each PVP, resulting in total nine predictors. We then use the ensemble to generate soft labels for 5,000 unlabeled data instances. The soft labels are then used, in conjunction with the labeled training data, to train a sequence classifier using RoBERTa-large. We observe in Figure~\ref{fig:conflict-patterns} that PET outperforms supervised learning on all training set sizes. This is true even for $|\trainset|=32$, where none of the predictors in the ensemble surpassed the performance of standard supervised learning.

Table~\ref{tab:pet-vs-supervised} provides a detailed comparison of PET with supervised learning. Recall that RoBERTa-large is both the supervised learner and the underlying MLM for PET. We observe that, for all training set sizes, PET is superior to supervised learning in all of the considered metrics. On average, when using PET instead of supervised learning, we observe an increase of 2.3\% in accuracy, 2.9\% in macro $F_1$ score, and 2.2\% in weighted $F_1$ score.

\begin{table}[!ht]
    \centering
    \caption{PET versus standard supervised learning.}
    \label{tab:pet-vs-supervised}
    \resizebox{\linewidth}{!}{
        \begin{tabular}{rcccccc}
        \toprule
              & \multicolumn{2}{c}{Accuracy} & \multicolumn{2}{c}{Macro $F_1$-score} & \multicolumn{2}{c}{Weighted $F_1$-score} \\
          \cmidrule(lr){2-3} \cmidrule(lr){4-5} \cmidrule(lr){6-7}
        $|\trainset|$  &  Supervised &  PET &  Supervised &  PET &  Supervised &  PET \\
        \midrule
        32     &              80.3 &          81.3 &                    62.8 &                64.8 &                       75.7 &                   77.1 \\
        64     &              78.3 &          81.4 &                    67.5 &                69.0 &                       77.1 &                   79.1 \\
        128    &              80.0 &          82.0 &                    70.0 &                71.7 &                       78.9 &                   80.3 \\
        256    &              81.6 &          85.4 &                    73.4 &                78.3 &                       80.9 &                   84.6 \\
        512    &              83.6 &          86.7 &                    75.7 &                81.0 &                       82.8 &                   86.3 \\
        1,024   &              86.6 &          87.4 &                    80.4 &                82.2 &                       86.1 &                   87.2 \\
        2,048   &              87.2 &          89.2 &                    81.4 &                84.8 &                       86.7 &                   89.0 \\
        \bottomrule
        \end{tabular}
}
\end{table}

Table~\ref{tab:pet-confusion} shows the confusion matrix for $|\trainset|=256$, where bolded terms indicate better performance. We observe that PET outperforms or matches the performance of supervised learning in every entry. As we previously noted, differentiating between conflict and duplicate appears to be a more difficult task as each label contains a similar vocabulary. In contrast, neutral statements are often easier to detect as indicated by these results. 

\begin{table}[!ht]
    \centering
    \caption{RoBERTa-large PLM / PET confusion matrix $\mathbf{|\trainset|=256}$.}
    \label{tab:pet-confusion}
    \begin{tabular}{lcccc}
        \toprule
         {} & \multicolumn{3}{c}{Predicted Label} & {} \\
        \cmidrule(lr){2-4}
        {} &   Conflict &  Duplicate &  Neutral & $F_1$-score \\
        True Label &  Sup. / PET & Sup. / PET & Sup. / PET & Sup. / PET \\
        \midrule
        Conflict  &  890 / \textbf{941} &   141 / \textbf{97} &     14 / \textbf{7} & 83.0 / \textbf{86.6} \\
        Duplicate &  199 / \textbf{179} &  112 / \textbf{136} &      4 / \textbf{0} & 39.5 / \textbf{49.6} \\
        Neutral   &     10 / \textbf{9} &      \textbf{0} / \textbf{0} &  630 / \textbf{631} & 97.9 / \textbf{98.7} \\
        \bottomrule
    \end{tabular}
\end{table}

\begin{figure*}[!ht]
    \centering
    \subfloat[Conflict recall.]{\includegraphics[width=0.33\textwidth]{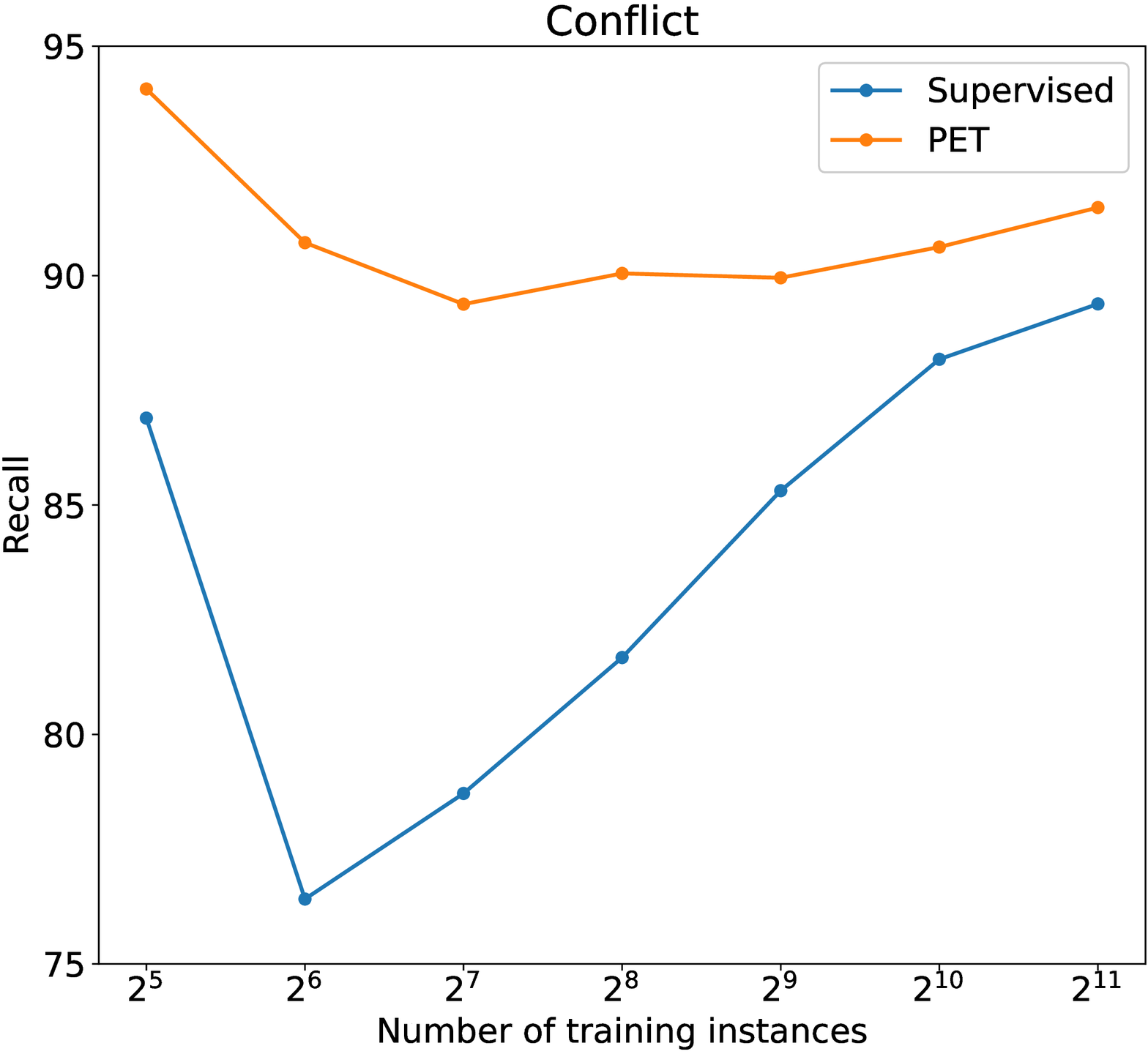}}
    \subfloat[Duplicate recall.]{\includegraphics[width=0.33\textwidth]{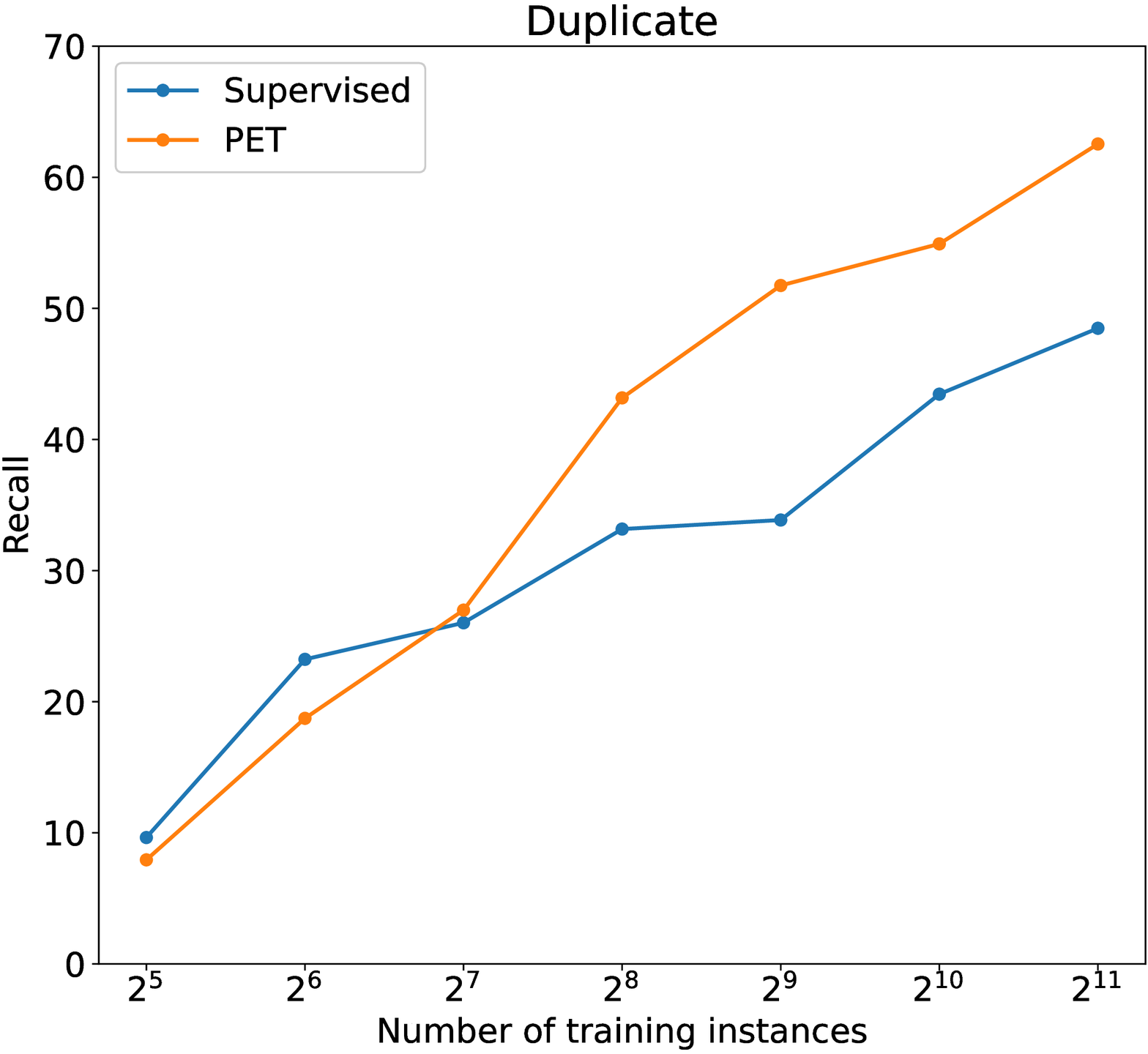}}
    \subfloat[Neutral recall.]{\includegraphics[width=0.33\textwidth]{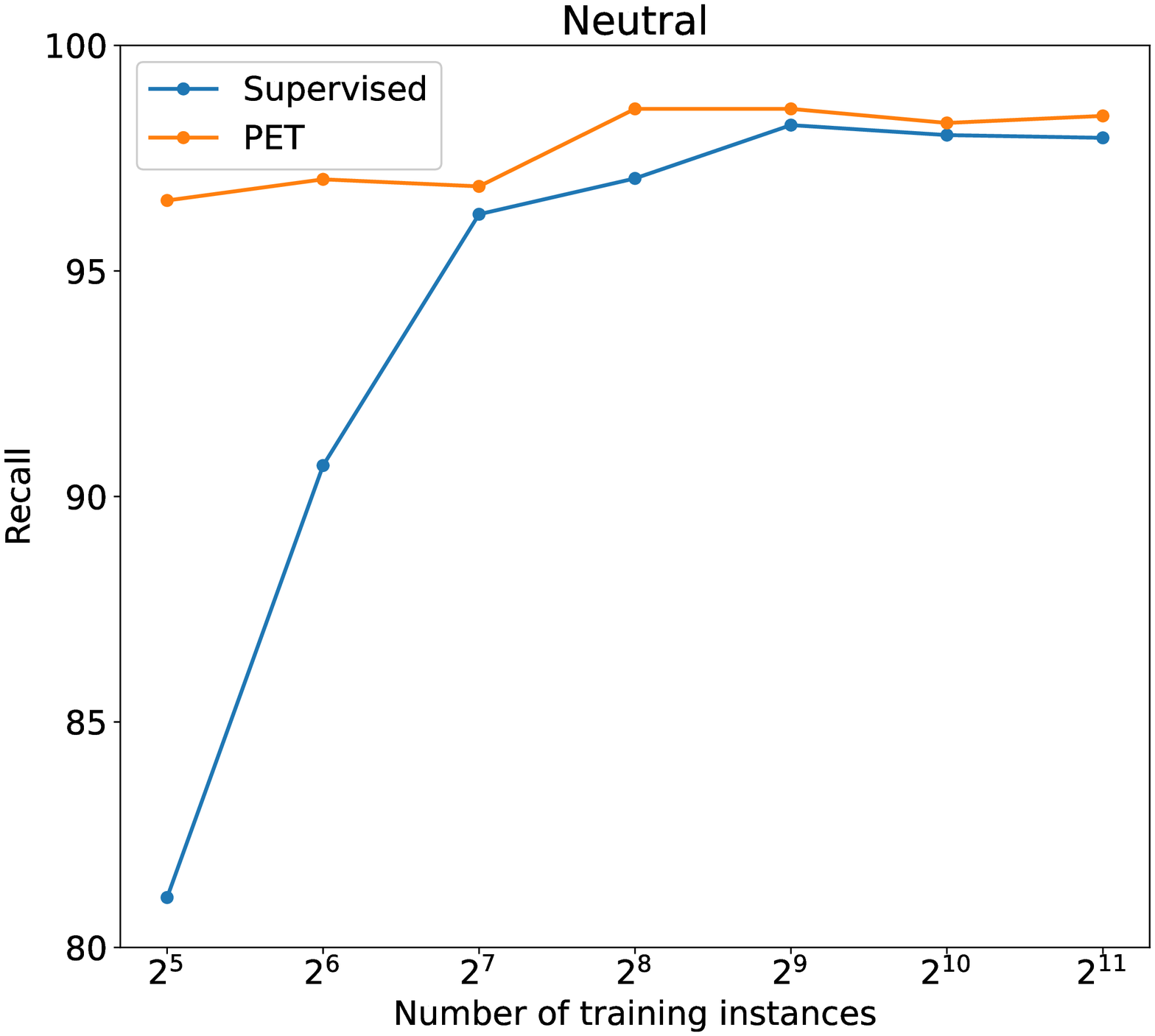}} \\*[0.8em]
    \subfloat[Conflict precision.]{\includegraphics[width=0.33\textwidth]{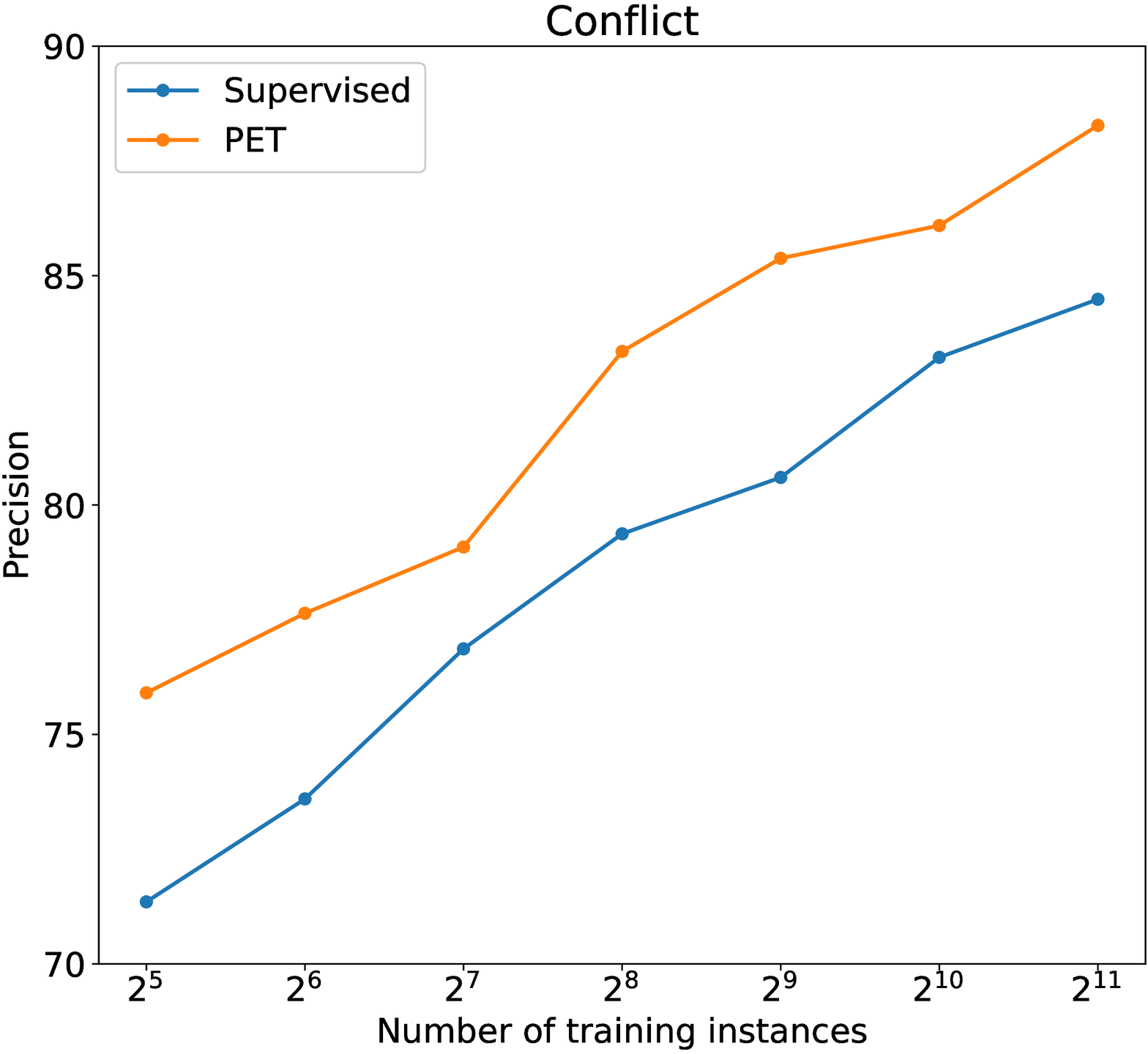}}
    \subfloat[Duplicate precision.]{\includegraphics[width=0.33\textwidth]{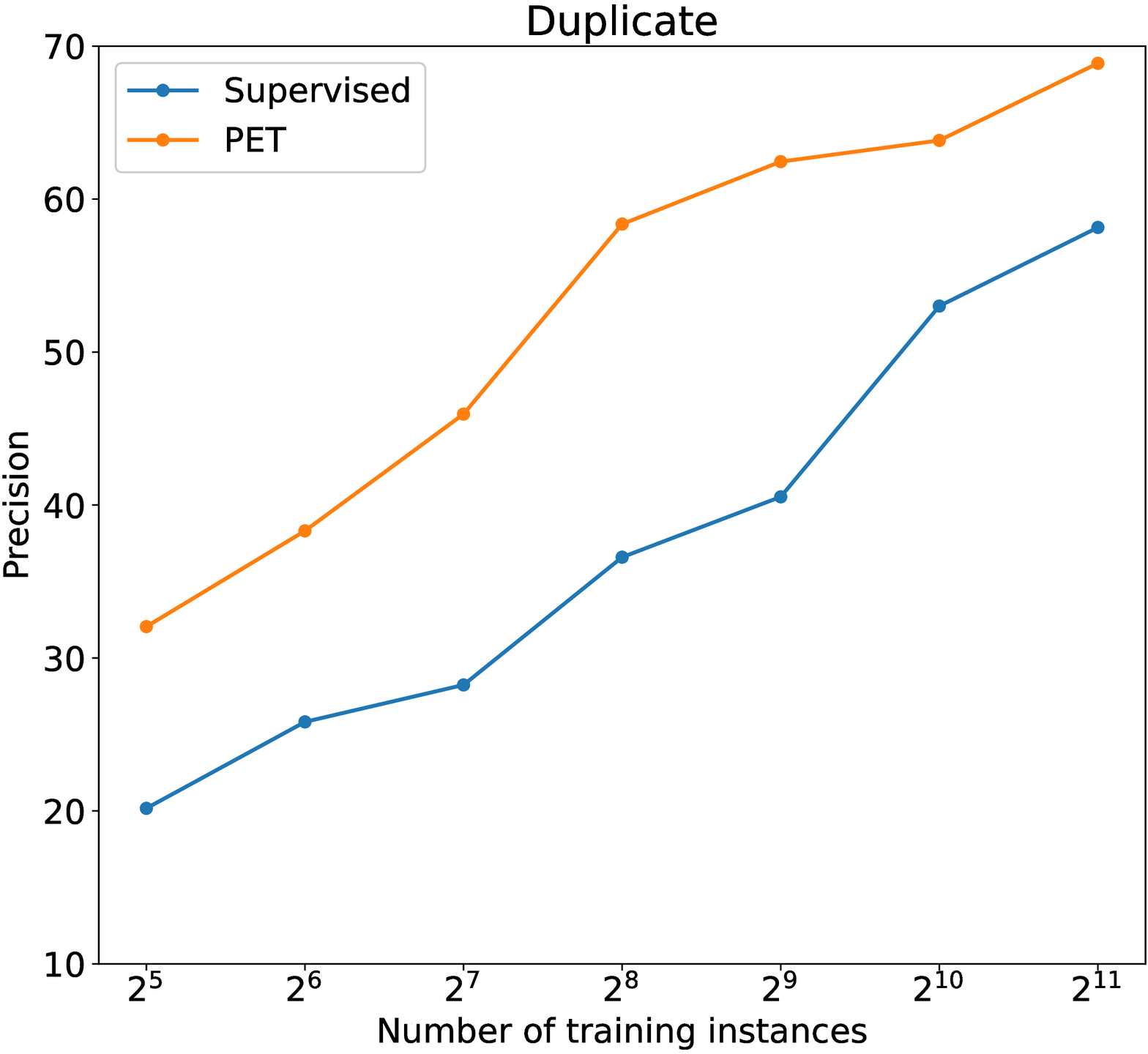}}
    \subfloat[Neutral precision.]{\includegraphics[width=0.33\textwidth]{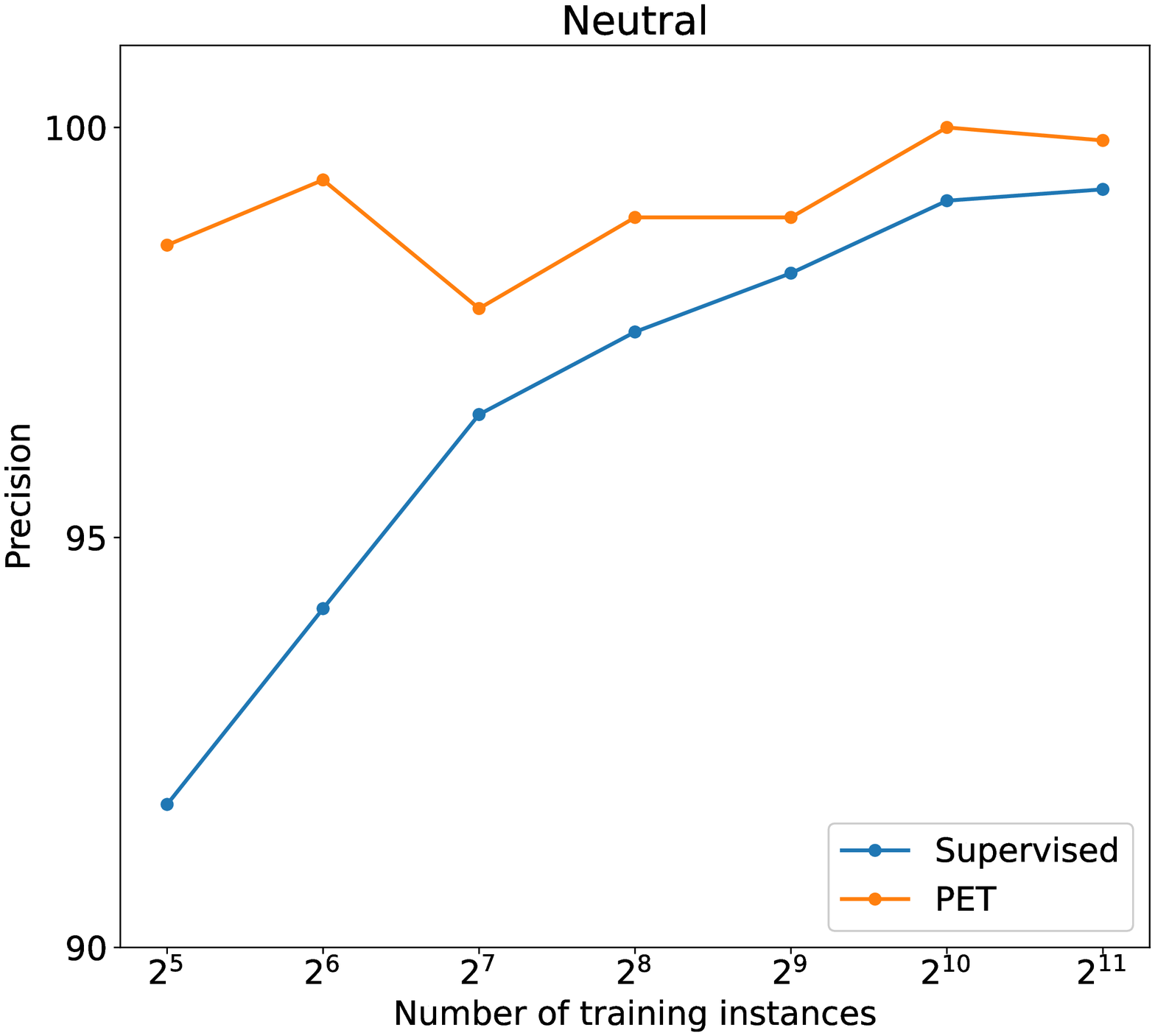}} \\
    \caption{Class-specific recall and precision values.}
    \label{fig:label-performance}
\end{figure*}

Figure~\ref{fig:label-performance} provides a detailed overview of the impact of PET versus supervised learning on a per-label basis. For the \textit{Conflict} label, traditional supervised learning with even 2,048 training samples is unable to match the recall of the PET algorithm with 32 training samples. This is despite the fact that the PET algorithm outperforms supervised learning in precision across all training set sizes. Additionally, precision for \textit{Conflict} using supervised learning requires 128 training samples to match that of PET with just 32 training instances.
In terms of recall, for small training set sizes both PET and supervised learning perform similarly, with supervised learning boasting a slight advantage for small training set sizes before PET widens the gap to nearly 10\% as $|\trainset|$ increases. However, we note that both models have a difficult time recalling pairings labeled \textit{Duplicate} when the training set is small. The precision of PET maintains a 10\% lead on supervised learning for all training set sizes when considering the \textit{Duplicate} label. Notably, the performance boost granted by PET for labeling duplicates is substantial even for large $|\trainset|$, with a boost of around 10\% in both precision and recall for the training set of size 2,048. In contrast, the gains for both \textit{Conflict} and \textit{Neutral} tend to subside for large $|\trainset|$ as they tend to approach their maximum values.

The performance gain in the neutral category is particularly noticeable. We observe that supervised learning requires 1,024 labeled examples in order to match the precision of PET with just 32 examples. Additionally, with access to just 32 labeled examples, PET achieves a precision of 98.6\% while recalling 96.6\% of examples. This precision margin is nearly perfect and can be conceivably used in a practical setting after training on just 32 labeled examples. This is a valuable result in practice as requirements marked \textit{Duplicate} or \textit{Conflict} both will generally have to be manually reviewed regardless. As a result, using PET with just 32 labeled examples can reliably identify pairings that must be reviewed with incredibly high recall and precision.

\section{Conclusion and Future Work}\label{sec:conclusion}
In this study, we demonstrated the viability of few-shot learning for conflict detection in SRS documents. We have shown that while pretrained language models such as BERT and its variants can be used in a few-shot setting, their performance is substantially worsened when access to data is limited and, furthermore, the relative performance of each model in a few-shot setting is considerably different from a data-rich setting. We also created our own cloze-style reformulations targeted toward conflict detection and compared them with general reformulations adopted from other tasks. Our results showed no considerable difference in performance, suggesting that adopting patterns and verbalizers from other tasks where possible may be sufficient for training a prompt-based learner. Finally, we provided a detailed comparison of PET with the best performing supervised learner and saw that the performance of PET in a few-shot setting was comparable to a supervised learner trained on considerably larger datasets.

We note that our study considers only a single conflict detection dataset, and while our results demonstrate that prompt-based learning is an effective candidate for few-shot learning in the conflict detection domain, future research into a broader array of datasets would be beneficial. Additionally, we note that both the choice of underlying language model for pattern-exploiting training and the selected prompts are both hyperparameters that are difficult to evaluate in a few-shot setting, and which can have a considerable impact on the performance of the final model.

This study offers many natural extensions into future work. Firstly, as noted, this research involves only a single data set. A more thorough investigation involving additional conflict detection tasks is necessary to establish PET as a viable, generic approach to conflict detection. Secondly, as the use of prompt-based learning for conflict detection is targeted towards few-shot learning, it is unlikely that validation data will be available to gauge the performance of PVPs ahead of time. Accordingly, while our results show that the patterns chosen have a substantial impact on performance, it is left for future research to determine why certain patterns perform better. Thirdly, while this study compares the performance of prompt-based learning with sequence classifiers, it does not compare PET to other few-shot learning approaches. Such a comparison would establish the advantage of using PET in practical settings.

\section*{Acknowledgements}
This research is supported by IBM CAS 1109.
The authors would like to thank IBM for providing support and feedback throughout this research. This research was also enabled in part by support provided by the Digital Research Alliance of Canada (\url{alliancecan.ca}). We would also like to thank the reviewers for their helpful comments and positive feedback.

\bibliographystyle{ACM-Reference-Format}
\bibliography{references}

\end{document}